\documentclass[aps,twocolumn]{revtex4}
 \usepackage[utf8x]{inputenc}
\usepackage{fontenc} 
\input{epsf}
\usepackage{epsfig}
\usepackage[]{graphicx} 
\usepackage{dcolumn} 
\usepackage{bm} 
\usepackage{slashed}
\usepackage{float}
\usepackage{amsmath,amssymb,amsfonts}
\usepackage{hyperref}
\usepackage{grffile}
\usepackage{fancybox}
\usepackage{setspace}
\usepackage{wrapfig}
\usepackage{lipsum}
\usepackage{color}

\usepackage[usenames,dvipsnames]{xcolor}
\newcommand{\sh}[1]{#1\hskip -6pt  / }

\newcommand{\up}[1]{u_\textbf{p}^s}
\newcommand{\vp}[1]{v_\textbf{p}^s} 
\graphicspath{ {figures/} }

\begin{document}

\title{A New Structure in the Deuteron}

\author{ Misak M. Sargsian and Frank Vera}
\affiliation{Florida International University, Miami, Florida 33199,USA
}
  
\date{\today}

\begin{abstract}
We demonstrate that a paradigm shift from considering the deuteron as a system of bound  proton and neutron 
to considering it as a pseudo-vector system in which we observe proton and neutron, results in a  possibility of probing a new 
``incomplete" P-state like  structure on the light-front (LF), at extremely large internal momenta, which can be achieved in 
high energy transfer electro-disintegration of the deuteron.
Investigating the deuteron on the light-front,  where the vacuum-fluctuations are suppressed,
we found that this new structure,  together with conventional 
S- and D- states,  is a leading order in transferred energy   of the reaction, thus 
it  is not suppressed on the light-front.
The incompleteness of the  observed P-state results in a violation of 
angular condition which can happen only if deuteron contains non-nucleonic structures 
such as $\Delta$$\Delta$, $N^*N$ or hidden color components. 
We demonstrate that experimentally verifiable signatures of ``incomplete" P-states are angular anisotropy of 
LF momentum  distribution of the nucleon in the deuteron as well as an enhancement of 
the tensor polarization strength beyond the  S- and D- wave predictions at large internal momenta in the deuteron. 
\end{abstract}
\maketitle

One of the outstanding issues of strong interaction physics is the understanding of the dynamics of 
transition between hadronic to quark-gluon phases of matter.
Such transitions at high temperature is relevant to the  evolution of the universe after the big bang 
and can be studied experimentally in heavy ion collisions. Transitions at low (near zero) temperatures  and high densities 
(``cold-dense" transitions)  
are relevant for superdense nuclear matter that can exists at the cores of neutron stars  and can set up 
the limits of matter density before it collapses to the black hole. However the direct  exploration of ``cold-dense" transitions is severely restricted.
 
Currently the accepted ways of investigating such transitions are; \newline
\indent (1) Studying nuclear medium modification of quark-gluon structure of  bound nucleons:
Such a modification was discovered in 1983 - by European Muon Collaboration\cite{EuropeanMuon:1983wih} - commonly referred to as EMC effect.
Few progresses were made in understanding of
 this phenomena  for past 40 years 
 (for reviews see \cite{Frankfurt:1988nt,Hen:2016kwk}), 
including the observation of the dependence of the effect on local nuclear density\cite{Seely:2009gt} and the important  role of 
short range nucleonic correlations in 
the  EMC effect for medium to large nuclei\cite{Weinstein:2010rt,CLAS:2019vsb}.
In all these cases the role of the hadronic to quark-gluon transition is not clearly understood.
\newline
\indent (2) Studying  the implications of the transition of baryonic matter to the quark matter in the cores of neutron stars.
The situation with the existence of quark matter in the cores of  Neutron Stars even more unclear than with the EMC effect.
With the observation of  unexpectedly large neutron star  masses\cite{Fonseca:2021wxt} ($\approx 2.08 M_{\odot}$) it was expected 
that if such stars would have 
radii, $R< 10$~km it will be indicative of large quark  matter component in their cores.
However observed radiuses for the large mass neutron stars 
are above  $R \ge 12$~km (e.g. Ref.\cite{Miller:2021qha}).

While  progress in advancing the studies of EMC effects is seen in performing  new generation of 
experiments  in which density of nuclear medium is controlled by tagging  a recoil nucleon which is 
in short range correlation  with the  probed nucleon (e.g. Ref.\cite{Melnitchouk:1996vp}).  The neutron star studies rely on improving  of 
detection techniques  that will allow to identify anomalously small size neutron stars.

In the present work we are suggesting a new  method of studying baryon-quark transition using  simplest known 
atomic nucleus,  the deuteron.

{\bf Deuteron on the Light Front~(LF):}
Our current  mindset about deuteron is fully non-relativistic,  within which,  the observation that  it has 
total spin, $J=1$  and positive parity, $P$,  together with the  relation that  for non-relativistic wave function, 
$P= (-1)^l$,  one concludes that the deuteron consists of S- and D- partial waves for proton-neutron system. 

However if  we are interested in deuteron structure at internal momenta comparable with the nucleon rest mass 
then nonrelativistic framework is not valid and  the problem is more fundamental, related to 
the description of a relativistic bound system. This situation is similar 
to the description of quark structure of nucleon in QCD in which case due to the small masses of u- and d- quarks the
vacuum fluctuations may overshadow the composite structure of the nucleon (see e.g. Ref.\cite{Feynman:1973xc}).

To discuss  relativistic structure of the deuteron on needs to identify the process in which 
the deuteron structure is probed. In our case we consider high-momentum transfer electrodisintegration process:
\vspace{-0.2cm}
\begin{equation}
e + d \rightarrow e^\prime + p + n
\label{reaction}
\vspace{-0.2cm}
\end{equation}
in which one of the nucleons are struck by the incoming probe and the spectator nucleon is probed with momenta comparable 
to nucleon masses.  If one can neglect (or remove) the effects related to final state interactions of two outgoing nucleons, 
then the above reaction at high $Q^2$, measures the probability of observing  proton and neutron in the deuteron 
at very large relative  momenta.  
In such a formulation the deuteron is not a composite system consisting of proton and neutron but it is 
a composite pseudo - vector ($J=1$, $P=+$)  ``particle" from which one extracts proton and neutron.
 How such a  proton and neutron are produced at   such extremal conditions is related 
to the dynamical  structure of LF deuteron wave function, which may include internal elastic $pn \rightarrow pn$ as well as inelastic 
$\Delta\Delta\to pn$, $N^*N\to pn$ or $N_cN_c\to pn$ transitions.  
Here, $\Delta$ and $N^*$ denote $\Delta$-isobar and $N^*$ resonances,  while 
$N_c$ is a  color octet baryonic state contributing to  the hidden-color component in the deuteron.

\vspace{-0.21cm}
The framework for calculation of reaction (\ref{reaction}) in relativistic domain  is LF
approach (e.g. Ref.\cite{Frankfurt:1981mk,Miller:2000kv,Brodsky:1997de,Frederico:1991vb,Vera:2021rnw,inprogress}) in which  one introduces 
LF deuteron wave function in the form:
\begin{equation}
\psi_{d}^{\lambda_d}(\alpha_i,p_\perp,\lambda_1\lambda_2) = - {\bar u(p_2,\lambda_2)\bar u(p_1,\lambda_1) \Gamma^\mu_{d} \chi_\mu^{\lambda_d}\over 
 {1 \over 2} ( m_d^2 - 4 {m_N^2 + p_\perp^2\over \alpha_i(2-\alpha_i)})\sqrt{2(2\pi)^3}}, 
\vspace{-0.3cm}
 \label{dwave_lf}
\end{equation}
where $\alpha_i = 2{p_{i+}\over p_{d+}}$, ($i=1,2$) and $\alpha_1+\alpha_2 = 2$ are LF momentum fractions of 
two nucleons outgoing from deuteron that has four-momentum $p_d^\mu$.\\
\indent Absorbing the energy denominator into the vertex function and using crossing symmetry one obtains:
\vspace{-0.19cm}
\begin{eqnarray}
&&\hspace{-0.2cm} \psi_{d}^{\mu}(\alpha_i,p_\perp,\lambda_1,\lambda_2) = 
 -\bar u(p_2,\lambda_2) \Gamma^\mu_{d}(k) {(i \gamma_2\gamma_0)\over \sqrt{2}} \bar u(p_1,\lambda_1)^T
  \nonumber \\ 
&&\hspace{-0.2cm}  = -\sum\limits_{\lambda_1^\prime}\bar u(p_1,\lambda_1) 
\Gamma^\mu_{d}\gamma_5 {\epsilon_{\lambda_1,\lambda_1^\prime}\over \sqrt{2}}u(p_1,\lambda^\prime_1),
\vspace{-2.cm}
\label{dwave_lf_b}
\end{eqnarray}
where $u(p,\lambda)$'s are LF bi-spinors of proton and neutron\cite{Lepage:1980fj} and 
$\epsilon_{i,j}$ is two dimensional Levi-Civita tensor, with $i,j=\pm 1$ helicity of nucleon. Since the deuteron is a pseudo-vector 
``particle",  due to $\gamma_5$ in Eq.(\ref{dwave_lf_b}), the vertex $\Gamma^\mu_d$ is a four-vector which we can construct
in a general form that explicitly satisfies time reversal, parity and charge conjugate symmetries. Noticing  that  
at the   $ d\to pn$ vertex  on the light-front 
the  "-"  ($p_- = E-p_z$) components 
of the four-momenta of the particles are not conserved, in addition to the four-momenta of two nucleons, $p_1^\mu$ and $p_2^\nu$  
one  has additional four-momentum:
\begin{equation}
\Delta^\mu \equiv p_1^\mu + p_2^\mu - p_d^\mu   \equiv (\Delta^-, \Delta^+, \Delta_{\perp}) = (\Delta^-, 0, 0),
\label{DeltaDefLF}
\end{equation}
where
\vspace{-0.4cm}
\begin{eqnarray}
\vspace{-0.4cm}
& & \hspace{-0.8cm} \Delta^- =  p_1^- + p_2^- - p_d^- = {m_N^2 + k^2_\perp\over p_1^+} +  {m_N^2 + k^2_\perp\over p_2^+} - {M_d^2\over p_d^+}  \nonumber \\
& & \hspace{-0.8cm}= {1\over p_d^+}\left[{4(m_N^2 + k^2_\perp)\over \alpha_1(2-\alpha_1)} - M_d^2\right] = 
{4\over p_d^+}\left[m_N^2 - {M_d^2\over 4} + k^2\right]. 
\vspace{-0.6cm}
\label{Delta-}
\end{eqnarray}
Here $k$ is the relative momentum 
in the $pn$ CM system defined as:
\begin{equation}
k = \sqrt{{m_N^2 + k_\perp^2\over \alpha_1(2-\alpha_1)} - m_N^2} \ \ \ \mbox{and} \ \ \ \alpha_1 = {E_k + k_z\over E_k},
\label{krel}
\end{equation}
where $E_k = {m^2 + k^2}$.
With $p_1^\mu$, $p_2^\mu$ and $\Delta^\mu$ 4-vectors the $\Gamma_d^\mu$ 4-vector function is constructed in the following form:
\begin{eqnarray}
\vspace{-0.4cm}
 & & \hspace{-0.2cm} \Gamma_d^{\mu}= \Gamma_{1}  \gamma^{\mu} +\Gamma_{2} \frac{(p_1-p_2)^{\mu}}{2m_N} + \Gamma_{3} \frac{\Delta^{\mu}}{2m_N}+
 \Gamma_{4} \frac{(p_1-p_2)^{\mu} \sh{\Delta}}{ 4m_N^2}     \nonumber \\
& &    +  i \Gamma_{5} \frac{1}{4 m_N^{3}} 
\gamma_{5} \epsilon^{\mu \nu \rho \gamma}(p_{d})_\nu (p_1-p_2)_{\rho} (\Delta)_\gamma 
 +   \Gamma_{6} \frac{ \Delta^{\mu} \sh{\Delta}}{4m_N^{2}},
\label{vertex}
\end{eqnarray}
where $\Gamma_i$,~($i=1,6$) are  scalar functions describing dynamics of  $pn$ component being observed in the deuteron.
 
\noindent{\bf High Energy Approximation:} 
For large $Q^2$ limit, LF momenta for reaction (\ref{reaction})  are chosen as follows:
 \vspace{-0.2cm}
 \begin{eqnarray}
&&\hspace{-0.2cm}p_d^\mu  \equiv  (p_{d}^-, p_{d}^+ ,p_{d\perp}) =  \left({Q^2\over x\sqrt{s}}\left[1 + {x\over \tau} - \sqrt{1+ {x^2\over \tau}}\right] \right., 
\nonumber \\  
&& \ \ \ \ \ \ \ \ \ \ \ \   \ \ \ \ \ \ \ \ \ \ \ \   \left.                          {Q^2\over x\sqrt{s}} \left[1 + {x\over \tau} +  \sqrt{1+ {x^2\over \tau}}\right], 0_\perp\right)    \nonumber \\
&&q^\mu \equiv  (q^{-},q^{+}, q_{\perp}) =  \left({Q^2\over x\sqrt{s}}\left[1 - x + \sqrt{1+ {x^2\over \tau}}\right], \right.
\nonumber \\  
&& \ \ \ \ \ \ \ \ \ \ \ \   \ \ \ \ \ \ \ \   \left.         {Q^2\over x\sqrt{s}}\left[1 - x - \sqrt{1+ {x^2\over \tau}}\right], 0_\perp\right),  
\label{refframeQ}
\end{eqnarray}
%
where $s = (q+p_d)^2$, $\tau={Q^2\over M_d^2}$ and $x = {Q^2\over M_dq_0}$, with $q_0$ being virtual photon energy in 
the deuteron rest frame.  The high energy nature of this process results in,  $p_d^+ \sim \sqrt{Q^2}\gg m_N$.
Then one observes in  Eq.(\ref{Delta-}) that the  $\Delta^-$ term is suppressed by the large $p_d^+$ factor.

Analyzing now the vertex function (\ref{vertex}) 
one observes that ${\Delta^-\over 2m_N}$ enters as a small 
parameter in the problem  in which $\Gamma_3$ and $\Gamma_4$ terms enter with the  order of ${\cal O}^1({\Delta^-\over 2m_N})$ 
while $\Gamma _6$ term enters as  ${\cal O}^2({\Delta^-\over 2m_N})$.  
Situation with $\Gamma_5$ term, is however  different; since for the covariant components:
$\Delta_+ = {1\over 2} \Delta^-$, $p_{d,-} = {1\over 2} p_d^+$,  
the  term with $\epsilon^{\mu + \perp -}$ is a
leading order (${\cal O}^0({\Delta^-\over 2m_N})$) due to the fact that  large $p_d^+$ factor is cancelled in $p_{d,-} \Delta_+ = 
{1\over 4} p_d^+ \Delta^-$ combination.

Keeping the leading, ${\cal O}^0(\Delta^-)$, terms in Eq.(\ref{vertex}) 
the LF deuteron wave function reduces to\cite{inprogress,Vera:2021rnw}:
 \vspace{-0.3cm}
\begin{eqnarray}
&& \hspace{-0.4cm}\psi_{d}^{\lambda_d}(\alpha_i,p_\perp) =  -\sum\limits_{\lambda_2,\lambda_1,\lambda_1^\prime}\bar u(p_2,\lambda_2) 
 \left\{ \Gamma_1\gamma^\mu +  \Gamma_2{(p_1-p_2)^\mu\over 2  m_N}   \right.   \nonumber \\
&& \hspace{-0.2cm}+\left. \sum\limits_{i=1}^{2} i\Gamma_5{1\over 8m^3_N}\epsilon^{\mu + i  -}p^+_{d} k_{i} \Delta^-\right\}  
 \gamma_5 {\epsilon_{\lambda_1,\lambda_i^\prime}\over \sqrt{2}}  u(p_1,\lambda^\prime_1)\chi_\mu^{\lambda_d},  
\label{dwave_lf3}
\end{eqnarray}
where 
$k_i = {(p_{1,i} - p_{2,i})\over 2}$, for $i=1,2$. The 
deuteron's polarization four-vector is chosen as:
\begin{equation}
\chi_\mu^{\lambda_d} = (\chi_0^{\lambda_d}, \chi_\perp^{\lambda_d} ,\chi_z^{\lambda_d}) = 
({p_{12} s_{d,z} \over M_{12}}, s_{s,\perp}, {E_{12} s_{d,z} \over M_{12}}),
\end{equation}
where ${\bf p_{12}} =  (p_{1_z} + p_{2,z}, 0_\perp)$,  $E_{12} = \sqrt{M_{12}^2 + p_{12}^2}$ and 
$M^2_{12}   =   s_{NN} = 4{(m_N^2+ k_\perp^2)\over \alpha_1(2-\alpha_1)}$.\\
\indent Since the wave function in Eq.(\ref{dwave_lf3})  is Lorentz boost invariant along the $z$ axis, it is convenient to 
calculate it in the deuteron CM frame obtained by boosting  with velocity $v = {{\bf p_{12}}\over E_{12}}$. Such a transformation 
results in\cite{inprogress}: 
 \vspace{-0.3cm} 
\begin{eqnarray}
& & \hspace{-0.3cm} \psi_{d}^{\lambda_d}(\alpha_i,k_\perp)   =     -\sum\limits_{\lambda_2,\lambda_1,\lambda_1^\prime}\bar u(-k,\lambda_2) 
 \left\{ \Gamma_1\gamma^\mu + 
\Gamma_2{{\tilde k}^\mu\over m_N}  + \right.    \nonumber \\
& &   \hspace{-0.3cm}  \left.    \sum\limits_{i=1}^{2}
 i\Gamma_5{1\over 8m^3_N}\epsilon^{\mu + i  -}p^{\prime +}_{d} k_{i} \Delta^{\prime-}\right\}   
  \gamma_5 {\epsilon_{\lambda_1,\lambda_i^\prime}\over \sqrt{2}} u(k,\lambda^\prime_1)s_\mu^{\lambda_d},   
\label{dwave_lf4}
\end{eqnarray}
where $\tilde k^\mu (0,k_z,k_\perp)$ with $k_\perp = p_{1\perp}$,
$k^2 = k_z^2 + k^2_\perp$ and  $E_{k} = 
{\sqrt{S_{NN}}\over 2}$  and  
$s_\mu^{\lambda_d} = (0,{\bf s^\lambda_d})$ in which:
 \vspace{-0.3cm} 
\begin{equation}
\hspace{-0.2cm} s_d^1 = - {1\over \sqrt{2}} (1,i,0), \  s_d^1 = {1\over \sqrt{2}} (1,-i,0) \   s_d^0 = (0,0,1). 
\end{equation}
In Eq.(\ref{dwave_lf4})  ``primed" variables correspond to the Lorentz boosts of respective 
unprimed quantities:
\begin{equation}
\hspace{-0.2cm}
p^{\prime +}_{d}   = \sqrt{s_{NN}}, \    
\Delta^{\prime-}  = {1\over \sqrt{s_{NN}}}\left[ {4(m_N^2 + k_\perp^2)\over\alpha_1(2-\alpha_1)}-M_d^2\right].
\vspace{-0.1cm}
\end{equation}
Since the term related to the 
$\Gamma_5$ is proportional to  ${4(m_N^2 + k_\perp^2)\over\alpha_1(2-\alpha_1)}-M_d^2$ which 
diminishes at small momenta,  only the $\Gamma_1$ and $\Gamma_2$ terms will contribute in nonrelativistic limit  defining 
the  $S$- and $D$- components of the deuteron.  Thus LF wave function in Eq.(\ref{dwave_lf4}) provides a smooth transition to 
the non-relativistic deuteron wave function.
This can be seen by expressing Eq.(\ref{dwave_lf4})  through two-component spinors:
\vspace{-0.2cm}
\begin{eqnarray}
 \psi_d^{\lambda_d}(\alpha_1,k_t,\lambda_1,\lambda_2)  =  
\sum\limits_{\lambda_1^\prime}\phi^\dagger_{\lambda_2} \sqrt{E_k}\left[  {U(k)\over \sqrt{4\pi}} {\bf \sigma s_d^{\lambda_d}}\right. - & & \nonumber \\
  - \left.    {W(k)\over \sqrt{4\pi}  \sqrt{2}}\left( { 3{\bf (\sigma k)(k s_d^\lambda)}\over k^2} - {\bf \sigma s_d^\lambda} \right) +\right. & &    \nonumber \\  
  \left.    
 (-1)^{1+\lambda_d\over 2} P(k)Y_{1}^{\lambda_d}(\theta,\phi)\delta^{1,\mid \lambda_d\mid}
  \right]
{\epsilon_{\lambda_1,\lambda_1^\prime}\over \sqrt{2}} \phi_{\lambda^\prime_1}. & &
\label{WF_LF}
\end{eqnarray}
Here the first two terms have explicit $S$- and $D$- structures were radial functions defined as: 
\begin{eqnarray}
& \hspace{-0.4cm} U(k)  = & {2\sqrt{4\pi} \sqrt{E_k}\over 3}\left[\Gamma_1(2+{m_N\over E_k}) + \Gamma_2{k^2\over m_N E_k}\right]\nonumber \\
& \hspace{-0.4cm} W(k)  = &  {2\sqrt{4\pi} \sqrt{2E_k}\over 3}\left[\Gamma_1(1-{m_N\over E_k}) -   \Gamma_2{k^2\over m_N E_k}\right].  
\label{radialwaves}
\end{eqnarray}
This relation is  known for $pn$-component deuteron wave function\cite{Frankfurt:1981mk,Carbonell:1995yi}, which allows to 
model  LF wave function through known radial $S$- and $D$- wave functions  estimated at LF relative momentum $k$ defined  in
Eq.(\ref{krel}).

However in addition to $S$-, $D$-  terms, our observation is that due to the $\Gamma_5$ term,  there is an additional leading contribution, 
which because of the relation  $Y^{\pm}_1(\theta,\phi) = \mp i\sqrt{3\over 4\pi}\sum\limits_{i=1}^{2}{ (k\times s_d^{\pm 1})_z\over k}$, has a 
$P$-wave like structure, where $P$- radial function  is defined as:
\begin{eqnarray}
&  \hspace{-0.4cm}P(k)  = &  \sqrt{4\pi}  {\Gamma_5(k) \sqrt{E_k}\over \sqrt{3}}{k^3\over m_N^3}.   
\label{Pradialwave}
\end{eqnarray}
It is worth emphasizing that this term  is purely relativistic in origin: as it follows from Eq.(\ref{Pradialwave})  it has  an extra ${k^2\over m_N^2}$ factor 
in addition to the ${k^{l=1}\over m_N}$ term characteristic to the radial $P$-wave. As a result one has a smooth transition to $S$- and $D$-states
in nonrelativistic limit.

The interesting feature of the above result which we will discuss in the next section, is that the $P$-wave is ``incomplete", that it  contributes only 
for $\lambda_d = \pm 1$ polarizations of the deuteron.

Closing this section we would like to mention that consideration of six  invariant vertex functions and contribution of $P$-radial waves in 
relativistic description of the deuteron were discussed earlier in the literature, see e.g. Refs.\cite{Buck:1979ff,Carbonell:1998rj}.  However,
to the best of our knowledge the observation that $\Gamma_5$ is a leading term on the light-front (while $\Gamma_{3,4,6}$ terms are suppressed)  
in high energy limit and it results in a non-complete $P$-wave contribution are  original results of the present work.
 

\noindent{\bf Light Front Density Matrix of the Deuteron:}Using Eq.(\ref{WF_LF}) one defines unpolarized deuteron light-front density matrix in the form\cite{Frankfurt:1981mk,Frankfurt:1988nt}:
\vspace{-0.3cm}
\begin{equation}
\vspace{-0.3cm}
\rho_d({\alpha,k_\perp}) = {n_d(k,k_\perp)\over 2-\alpha},
\label{rho_unpol}
\end{equation}
where LF momentum distribution is expressed through the radial wave functions as follows:
\vspace{-0.3cm}
\begin{eqnarray}
\vspace{-0.4cm}
n_d(k,k_\perp)  &  = &  {1\over 3}\sum\limits_{\lambda_d=-1}^{1}\mid \psi_d^{\lambda_d}(\alpha,k_\perp)\mid^2 = \nonumber \\
& = & {1\over 4\pi} \left( U(k)^2 +  W(k)^2 + {k_\perp^2\over k^2} P^2(k)\right).
\label{momdist}
\end{eqnarray}
The LF density matrix satisfies baryonic and momentum sum rules as follows:
\begin{equation}
\int \rho_d({\alpha,k_\perp}) {d\alpha\over \alpha} = 1 \ \ \ \mbox{and} \ \ \ \int\alpha \rho_d({\alpha,k_\perp}) {d\alpha\over \alpha} = 1.
\label{sumrules}
\end{equation}
From the above,  the normalization condition for the radial wave functions is:
\vspace{-0.3cm}
\begin{equation}
\int \left(U(k)^2 +  W(k)^2 + {2\over 3} P^2(k)\right)k^2 dk = 1.
\label{norm}
\end{equation}

\noindent{\bf The $\Gamma_5$-Term and Non-Nucleonic Component in the Deuteron:}The unusual result of 
Eq.(\ref{WF_LF}) is that the $P$-wave like  term  enters only for deuteron polarizations, $\lambda_d = \pm 1$.
The later is the reason that momentum distribution in Eq.(\ref{momdist})  depends  explicitly on the transverse component of 
the relative momentum on the light front. 
Such a behavior is impossible for non-relativistic quantum mechanics of the 
deuteron since in this case the potential of the interaction is real (no inelasticities) and the solution of Lippmann-Schwinger equation 
for partial S- and D-waves satisfies ``angular condition", according to which the momentum distribution in unpolarized deuteron depends on 
the magnitude of relative momentum only.  Our result do not contradict  the properties of non-relativistic deuteron wave 
function since as it was discussed earlier according to Eq.(\ref{Pradialwave}) the P-wave is purely relativistic in its nature. 
On the other hand, in the relativistic 
domain the definition of the interaction  potential is not straightforward to allow to use quantum-mechanical arguments
in claiming  that momentum distribution in Eq.(\ref{momdist}) should satisfy the angular condition (i.e. depends on  magnitude of $k$ only). 

For the relativistic domain, on the light-front, the analogue of Lippmann-Schwinger equation is the Weinberg type 
equation\cite{Weinberg:1966jm} using which for $NN$ scattering amplitude, in which only nucleonic degrees are considered,
in the CM of the NN system, one obtains\cite{Frois:1991wc}:
\begin{eqnarray}
&&T_{NN}(\alpha_i,k_{i\perp},\alpha_f,k_{f,\perp}) \equiv T_{NN}(k_{i,z},k_{i\perp},k_{f,z},k_{f,\perp}) =\nonumber \\
& & V(k_{i,z},k_{i\perp},k_{f,z},k_{f,\perp})  + \int V(k_{i,z},k_{i\perp},k_{m,z},k_{m,\perp})\times   \nonumber \\
&  & 
\times{d^3 k_m\over (2\pi)^3 \sqrt{m^2 + k_m^2}}{T_{NN}(k_{m,z},k_{m\perp},k_{f,z},k_{f,\perp})\over 4(k_m^2 - k_f^2)},
\label{TNN}
\end{eqnarray} 
where ``i", ``m" and ``f", subscripts correspond to initial, intermediate and final $NN$ states respectively and momenta $k_{i,m,f}$ 
are defined similar to Eq.(\ref{krel}). 
The realization of the angular condition for relativistic case will require that light-front potential to satisfy a condition
\begin{equation}
V(k_{i,z},k_{i\perp},k_{m,z},k_{m,\perp}) = V(\vec k_i^2, (\vec k_m-\vec k_i)^2).
\label{Vangcond}
\end{equation}
Such a  conditions is obvious for on-shell limit, since the Lorentz invariance of the $T_{NN}$ amplitude requires:
\begin{equation}
T_{NN}^{on \ shell}(k_{i,z},k_{i\perp},k_{m,z},k_{m,\perp}) = T^{on \ shell}_{NN}(\vec k_i^2, (\vec k_m-\vec k_i)^2)
\label{Tangcond}
\end{equation}
and existence of the Born term in Eq.(\ref{TNN}) indicates that  the potential $V$ satisfies 
the same condition  in the  on-shell limit.   

For the off-shell potential the  angular condition  is not obvious. However  in Ref.\cite{Frankfurt:1988nt,FSMS90,Frois:1991wc} 
it was shown  that requirements of potential $V$ satisfying  angular condition in the on-shell limit and 
that it  can be constructed through the series of elastic $pn$ scatterings  result  in a potential which is analytic function of angular momentum.
Then with the minimal assumption that the potential, analytically continued to the complex angular momentum space, does not diverge 
exponentially,  it was shown that  $V$ and the $T_{NN}$ functions satisfy angular condition (Eqs.(\ref{Vangcond},\ref{Tangcond})) in general.
Then, using the same potential to calculate LF deuteron wave function will result in a momentum distribution that will depend on 
the magnitude of the relative $pn$ momentum only. This observation
requires the restriction by the $pn$ component only in the deuteron.  

Inclusion of the inelastic transitions will completely change the LF equation  for the $pn$ scattering. 
For  example, contribution of  $N^*N$ transition to the elastic $NN$ scattering:
\vspace{-0.3cm}
\begin{eqnarray}
&&\hspace{-0.4cm}T_{NN}(k_{i,z},k_{i\perp},k_{f,z},k_{f,\perp}) =  \int V_{NN^*}(k_{i,z},k_{i\perp},k_{m,z},k_{m,\perp})  \nonumber \\
& & 
\times{d^3 k_m\over (2\pi)^3 \sqrt{m^2 + k_m^2}} {T_{N^*N}(k_{m,z},k_{m\perp},k_{f,z},k_{f,\perp})\over 4(k_m^2 - k_f^2+m^2_{N*} - m_N^2)},
\label{TNN*}
\end{eqnarray} 
will not require the condition of Eq.(\ref{Vangcond}) with the transition potential having also an imaginary component.  Eq.(\ref{TNN*}
can not 
be described with any combination of elastic $NN$ interaction potentials that satisfies the angular condition.
The same will be true also  for $\Delta\Delta\to NN$  and   $N_c,N_c\to NN$ transitions.
This indicates that if the $\Gamma_5$ term is not zero then it should originate from non-nucleonic component in the 
deuteron.

\noindent{\bf Estimate of the Possible Effects:} Our prediction is that the observation of  anisotropic LF momentum distribution  depending on 
the center of mass $k$ and $k_\perp$ separately will indicate the presence of non-nucleonic component in the deuteron. 
Since this effect is due to  the $P$-wave  like structure,  (originating from $\Gamma_5$ term) which has an extra ${k^2\over m^2_N}$ 
factor (Eq.(\ref{radialwaves}))  compared to $S$- and $D$- radial waves, one expects it to become important at 
$k> m_N$.  

\begin{figure}[h]
\includegraphics[width=8.2cm,height=5.2cm]{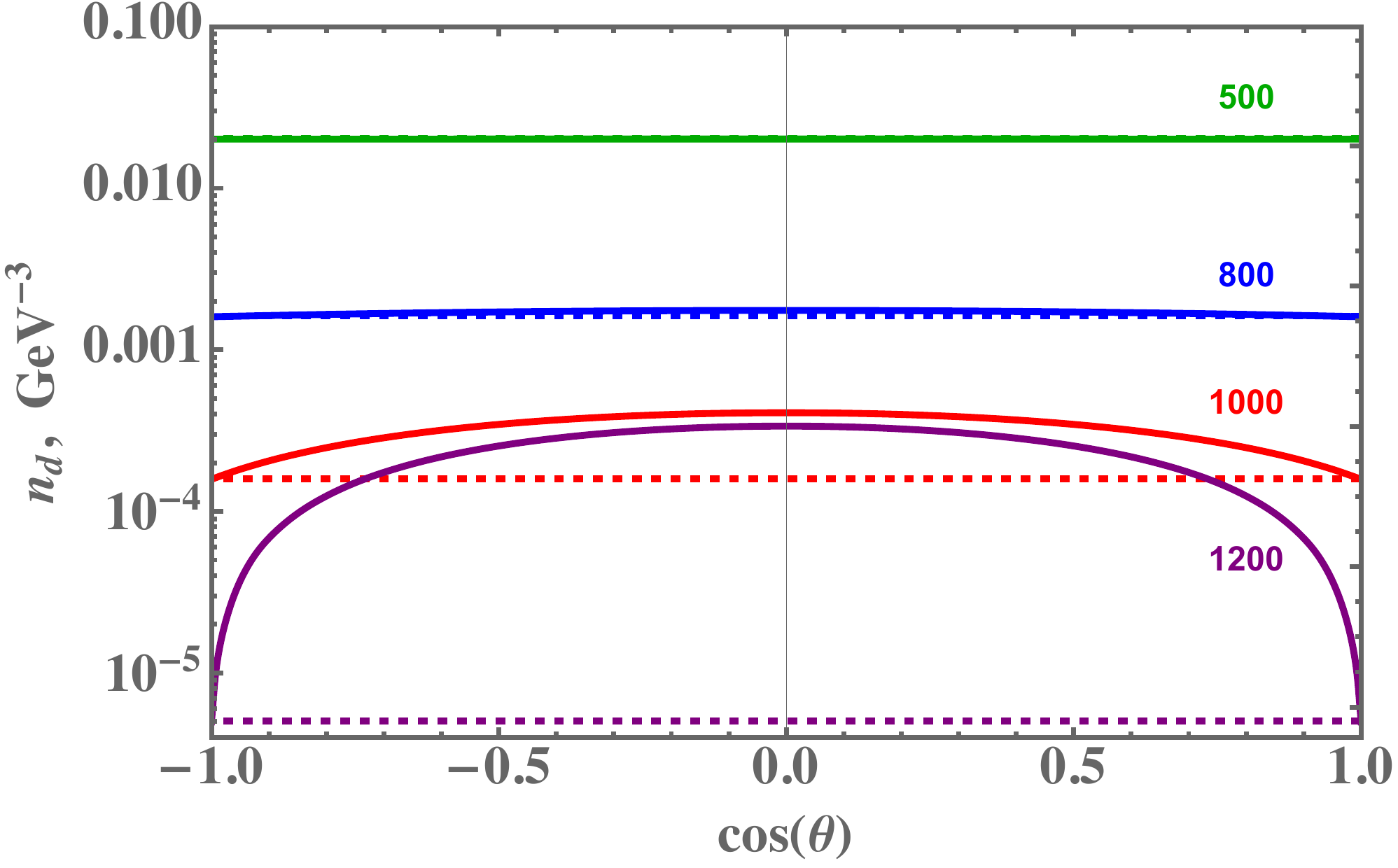}
\vspace{-0.5cm}
\caption{LF momentum distribution of the deuteron as a function of $\cos{\theta}$, for different values of $k$.
 Dashed lines - deuteron with $pn$ component only, solid lines - with $P$-wave like component included.}
\label{mbfigures1}
\end{figure}

To give qualitative estimate of the possible effects we evaluate $\Gamma_5$ vertex function assuming  two color-octet baryon transition to 
the $pn$ system ($N_cN_c\to pn$) through one-gluon exchange,  parameterizing it in the dipole form ${A\over  (1 + {k^2\over 0.71})^2}$. The parameter $A$ is estimated by assuming 1\% contribution to the total normalization from the $P$ wave in Eq.(\ref{norm}). The latter is consistent 
with the experimental estimation in Ref.\cite{deltadelta}  of 0.7\%. In Fig.\ref{mbfigures1} we consider dependence of the momentum 
distribution of Eq.(\ref{momdist}) as a function of $\cos{\theta} = {(\alpha-1)E_k\over k}$ for different values of $k$. Notice that if momentum distribution is 
generated by $pn$ component only, the angular condition is satisfied, and  no dependence should be observed.  
 
\begin{figure}[h]
\includegraphics[scale=0.36]{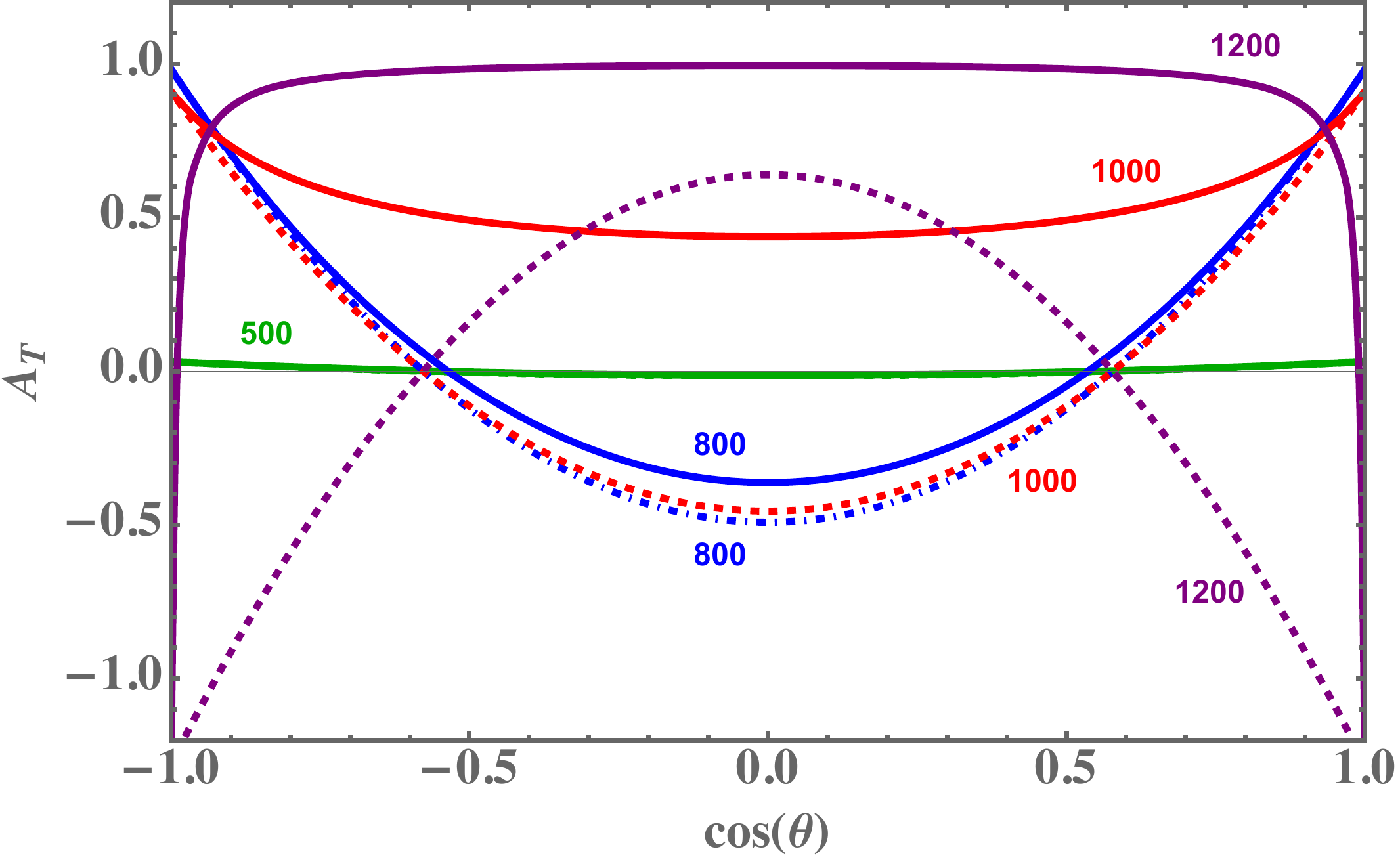}
\vspace{-0.5cm}
\caption{Tensor asymmetry as a function of  $\cos{\theta}$ for different $k$.  Dashed lines - deuteron with $pn$ component only,
solid lines - with $P$ component included.}
\label{mbfigures2}
\end{figure}

As the figure shows on may expect  measurable angular dependence  at $k\gtrsim 1$~GeV/C which is consistent with the expectation that inelastic transition in the deuteron corresponding to the non-nucleonic  components takes place at $k\gtrsim 800$~MeV/c.  Additionally due to  the fact that 
the $P$-component  contributes only for $\lambda_d = \pm 1$ polarization of the deuteron (Eq.(\ref{WF_LF}))  one expects enhanced effect 
in the asymmetry from scattering off the tensor polarized deuteron:
\begin{equation}
A_T = {n_d^{\lambda_d = 1}(k,k_\perp) +  n_d^{\lambda_d = -1}(k,k_\perp) - 2  n_d^{\lambda_d = 0}(k,k_\perp)\over  n_d(k, k_\perp)}.
\end{equation}
As Fig.\ref{mbfigures2} shows the presence of non-nucleonic component will 
be visible   already at $k\approx 800$~MeV/c resulting in a qualitative difference  
in asymmetry at larger momenta as compared with the asymmetry predicted  by deuteron wave function with $pn$-component only.
 
{\bf Outlook on Experimental Verification of the Effect:} Prediction that non-nucleonic component in the deuteron wave function 
may result in angular dependence of LF-momentum distribution can be verified at CM momenta $k \gtrsim 1$~GeV/c.  This seems 
incredibly large momenta to be measured in  experiment. However the first such measurement at high $Q^2$ disintegration of the deuteron is already performed at Jefferson Lab\cite{HallC:2020kdm} reaching $k\sim 1$~GeV/c.  It is intriguing that the results of this  measurement qualitatively disagree with  predictions based on conventional deuteron wave function once $k\gtrsim 800$~MeV/C.  The planned new measurement\cite{Boeglin:2014aca} will significantly improve the quality of the data allowing possible verification of the effects discussed in 
this work.  It is worth mentioning that the analysis of the experiment will require careful account for competing nuclear effects such as final state 
interaction for which there is a significant  theoretical and experimental progress during the last decade\cite{Sargsian:2009hf,Boeglin:2015cha}. 
If the experiment will not find the angular dependence in the momentum distribution this will allow to set a new limit on 
the dominance of $pn$ component at instantaneous high nuclear densities that corresponds to $\sim 1$~GeV/c internal momentum 
in the deuteron.  If, however,  the angular dependence is found, it will motivate theoretical modeling of non-nucleonic components in 
the deuteron, such as $\Delta\Delta$, $N^*N$ or hidden-color $N_cN_c$ that can reproduce the observed result. In both cases results of such 
studies will  advance the understanding of the dynamics of high density nuclear matter and the relevance of the quark-hadron transition.
Possibility of studies of tensor asymmetries will significantly complement  above studies, however feasibility of such experiments currently 
is not clear.

 {\bf Acknowledgments:}
We are thankful to Drs. Leonid Frankfurt and Mark Strikman for useful discussions. 
This work is supported by the U.S. DOE Office of Nuclear Physics grant DE-FG02-01ER41172.



\end{document}